\documentclass[twocolumn,secnumarabic,showpacs,amssymb,amsmath,nobibnotes, aps, prl]{revtex4}
\usepackage{graphics}
\usepackage{graphicx}
\usepackage {latexsym}
\begin{document}
\draft

\title{Synchronizing Huygens's clocks}

\author{Rui Dil\~ao}
\email{rui@sd.ist.utl.pt}
\affiliation{NonLinear Dynamics Group, Instituto Superior 
T\'ecnico, Av. Rovisco Pais, 1049-001 Lisbon, Portugal
}
\date{\today}

\begin{abstract}
We introduce an interaction mechanism between oscillators leading to exact anti-phase and in-phase synchronization. This mechanism is applied to the coupling between two nonlinear oscillators with a limit cycle in phase space, leading to a simple justification of the anti-phase synchronization observed in the Huygens's pendulum clocks experiment.  If the two coupled nonlinear  oscillators reach the anti-phase or the in-phase synchronized oscillatory state, the period  of oscillation is different from the eigen-periods of the uncoupled oscillators.  
\end{abstract}  


\maketitle

In  1665, Christiaan Huygens reported the observation of the synchronization of two pendulum clocks closely hanged on the wall of his workshop, \cite[pp. 357-361]{1}. After synchronization, the clocks swung in exactly the same frequency and $180^o$ out of phase. 
Huygens also noted that if the two clocks were hanged in such a way that the planes of oscillation of the two pendulums were mutually perpendicular, then  synchronization didn't occur. 
Huygens justified the observed synchronization phenomena by the ``sympathy that cannot be caused by anything other than the imperceptible stirring of the air due to the motion of the pendulum'', \cite{1}. 

Recently, Bennett {\it et al.}, \cite{2}, built an experimental device   consisting of two interacting pendulum clocks hanged on a heavy support, and  this support was mounted on a low-friction wheeled cart. This device  moves by the action of the tensions due to the swing of the two pendulums, and the interaction between the two clocks is caused by the mobility of the heavy base of the clocks. 
With this device, the anti-phase synchronization mode is reached when the difference between the natural or eigen-frequencies of the two clocks is less than $0.0009$~Hz. If the difference between these frequencies is larger than $0.0045$~Hz, the two clocks don't  synchronize, running ``uncoupled'' or in a state of beating death, \cite{2}. This situation is unsatisfactory when compared with the observations of Huygens. For example, a difference of order of $\Delta \omega=0.0009$~Hz for the two pendulum eigen-frequencies  corresponds to a difference in the lengths of the pendulum rods of the order of  $\Delta \ell = \sqrt{g}\ell^{3/2}\Delta \omega /4\pi$, which gives, for $\ell =1$~m and $g=9.8$~ms$^{-2}$,  $\Delta \ell=4$~mm, and  for  $\ell =0.178$~m (the length of the pendulum rods used by Huygens, \cite{1}), $\Delta \ell=0.02$~mm, a precision that Huygens certainly couldn't achieve. 
According to Bennett {\it et al.} \cite[p. 578]{2},   Huygens's results depended on both talent and luck.

Another experimental model mimicking the Huygens's clocks system,
consists of two pendulums whose suspension rods are connected by a weak string, and one of the two pendulums is driven by an external rotor, \cite{3} and \cite{4}.
 In this system, the in-phase synchronization is approximately achieved with a small phase shift, and  the experimental measurements and the model analysis both agree. 
The numerical results of Fradkov and Andrievsky for this device, \cite{4}, show simultaneous and  approximate in-phase and anti-phase synchronization, tuned by different initial conditions.
In another experimental device made of two rotors controlled by external torques  (\cite{5, 6}), Andrievsky {\it et al.}, \cite{5},
reported approximate anti-phase and in-phase synchronization of the
two  oscillators. In this experiment, the synchronization  parameter is the stiffness of a string connecting the two rotors. 

In  these experimental systems, there is no clear evidence
of what mechanism is in the origin of the anti-phase synchronization, as described by Huygens. In general, it is believed that if the pendulums have slightly different periods, the two oscillators may not synchronize,
\cite{2, 4}. These experimental studies seem to corroborate this conclusion. However, as this special type of collective rhythmicity occurs in biological systems and several other natural phenomena, \cite{7}, where individual  periods are different, it is important to derive and to understand the interaction mechanisms leading to exact synchrony.

In this paper, we introduce an interaction mechanism between oscillators leading to exact anti-phase and in-phase synchronization. The coupling between the oscillators is derived by modeling  
explicitly the physical processes involved in the interaction.
The oscillators under analysis can be simple harmonic oscillators, pendulums, or nonlinear oscillators with a limit cycle in phase space. 

In the Huygens two pendulum clocks system,
the pendulums are hanged in a common support, and  the only possible interaction between them is due to the tension forces generated by the oscillatory motion of the two pendulums.  These tension forces propagate  through the common support, that we consider to be elastic.  The role of the tension forces in the interaction  is corroborated by the Huygens's finding that when "the 
clockfaces were facing each other", \cite[p. 359]{1}, or
the planes of oscillation of the two pendulums are mutually perpendicular,
no synchronization is observed. In fact, the components of the tension forces generated by the motion of the pendulums are in the plane of motion of the pendulums. 

\begin{figure}
 \includegraphics[width=0.9 \hsize]{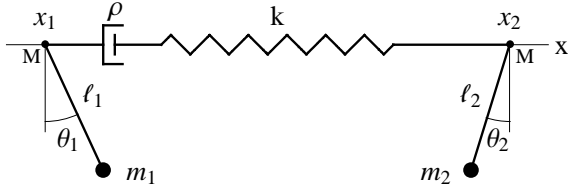}
 \caption{Model to analyze the synchronization of the Huygens's two-pendulum clocks system. The two pendulums are a representation of  two nonlinear oscillators. The interaction between the pendulums is done by the tension forces  at the attachment points, and they actuate through an elastic and resistive media. Each attachment points is considered to have mass $M$.}
\label{fig1} 
\end{figure}

To model the Huygens's experiment, we consider the geometric arrangement of Fig.~\ref{fig1}, where the two pendulums have masses $m_1$  and $m_2$, and lengths $\ell_1$  and  $\ell_2$, respectively. 
The pendulums are considered connected by a massless string with stiffness constant  $k$. The perturbations that propagate along the string are damped, and the damping force is proportional to the velocity of the attachment points of the string,  with damping constant  $\rho$. The string and the damping of the attachment points simulate  the elasticity
and the resistivity of the common support of the pendulums. 

We also assume that the attachment points of the pendulums have equal masses $M$, and their deviations from the rest positions  are measured by horizontal coordinates $x_1$  and  $x_2$, respectively.   As we shall see below, the introduction of the mass constant $M$ is necessary to obtain explicitly the equations of motion.

The system of Fig.~\ref{fig1}, considered without the damping forces, is described by the four degrees of freedom Lagrangian,
\begin{equation}
\begin{array}{ll}\displaystyle
 L  =&\frac{1}{2}m_1 (\ell_1^2 \dot \theta_1^2  + \dot x_1^2  + 2\ell_1 \dot x_1 \dot \theta_1 \cos \theta_1 )+ m_1 g\ell_1 \cos \theta_1   \\ [2pt]
&+ \frac{1}{2}m_2 (\ell_2^2 \dot \theta_2^2  + \dot x_2^2  + 2\ell_2 \dot x_2 \dot \theta_2 \cos \theta_2 ) + m_2 g\ell_2 \cos \theta_2\\ [2pt]
&+ \frac{1}{2}M(\dot x_1^2  + \dot x_2^2 ) - \frac{1}{2}k(x_2  - x_1 )^2 
 \end{array}
\label{eq1}
\end{equation}
where  $\theta_1$ and $\theta_2$ are the angular coordinates of the two pendulums,   $g$ is the acceleration due to the gravity force, and the last two terms describe the interaction between the two pendulums.  From (\ref{eq1}), the Lagrange equations of motion of the system of Fig.\ref{fig1} are, 
\begin{equation}
\begin{array}{lcl}\displaystyle
 m_1 \ell_1 \ddot \theta_1  + f_1 (\theta_1 ,\dot \theta_1 ) + 
m_1 g\sin \theta_1  &=&  - m_1 \ddot x_1 \cos \theta_1   \\ 
 m_2 \ell_2 \ddot \theta_2  + f_2 (\theta_2 ,\dot \theta_2 ) + 
m_2 g\sin \theta_2  &=&  - m_2 \ddot x_2 \cos \theta_2   \\ 
 (M + m_1 )\ddot x_1  + 2\rho \dot x_1  + m_1 \ell_1 \ddot 
\theta_1 \cos \theta_1  &=& m_1 \ell_1 \dot \theta_1^2 \sin \theta_1  \\ & &+ k(x_2  - x_1 ) \\ 
 (M + m_2 )\ddot x_2  + 2\rho \dot x_2  + m_2 \ell _2 \ddot \theta_2 \cos \theta_2  &=& m_2 \ell_2 \dot \theta_2^2 \sin \theta_2  \\ & &- k(x_2  - x_1 ) \\ 
 \end{array}
\label{eq2}
\end{equation}
where we have added the dissipative terms implicit in the interaction model of Fig.~\ref{fig1},
$\rho$  is the damping constant of the attachment points, and the functions  $f_1 (\theta_1 ,\dot \theta_1 )$ and $f_2 (\theta_2 ,\dot \theta_2 )$  describe the escaping mechanism
of the clocks. The terms in $\rho$, $f_1$ and $f_2$ are dissipative terms, not contained in the Lagrangian function (\ref{eq1}). 

The system of  equations (\ref{eq2}) implicitly defines a system of ordinary differential equations. If  $M>0$,  $\ell_1>0$,  $m_1>0$, $\ell_2>0$  and  $m_2>0$, the   system 
of equations (\ref{eq2}) can be solved algebraically in order to the higher order derivatives. Solving the system of  equations (\ref{eq2}) in order to the higher order derivatives, and introducing the assumption of small amplitude of oscillations, we obtain,
\begin{equation}
\begin{array}{lcl}
 m_1 \ell _1 \ddot \theta _1  + f_1 (\theta _1 ,\dot \theta _1 ) + m_1 g\theta _1  &=&  - m_1 \ddot x_1  \\ 
 m_2 \ell _2 \ddot \theta _2  + f_2 (\theta _2 ,\dot \theta _2 ) + m_2 g\theta _2  &=&  - m_2 \ddot x_2  \\ 
 M\ddot x_1  - f_1 (\theta _1 ,\dot \theta _1 ) + 2\rho \dot x_1  - m_1 g\theta _1  &=& k(x_2  - x_1 ) \\ 
 M\ddot x_2  - f_2 (\theta _2 ,\dot \theta _2 ) + 2\rho \dot x_2  - m_2 g\theta _2  &=&  - k(x_2  - x_1 ) \\ 
 \end{array}
\label{eq3}
\end{equation}

In the particular case of the pendulum clocks, we consider that its dynamics is well described by a nonlinear oscillator with a limit cycle in phase space, \cite{8}. To simplify,
we assume that the individual dynamics of each oscillator is described by the second order equation,
\begin{equation}
m \ell \ddot \theta  + f(\theta;\lambda ,{\tilde \theta})\dot \theta  + mg \theta  = 0
\label{eq4}
\end{equation}
where,
\begin{equation}
f(\theta; \lambda ,{\tilde \theta}) = \left\{ 
   \begin{array}{rcl}
-2\lambda   &\ \hbox{if}\ & |\theta | < \tilde \theta  \\ 
2\lambda     &\hbox{if}\ & |\theta | \ge \tilde \theta 
\end{array} 
\right.  
\label{eq5}
\end{equation}
and  $\lambda$ and $\tilde \theta $ are  positive  constants.  
A simple qualitative analysis, shows that the second order differential equation (\ref{eq4}) has a unique limit cycle in phase space, \cite{9}. 
If $\tilde \theta $ is small,  the small amplitude approximation used in the derivation of (\ref{eq3}) is still valid, and the mean radius of the limit cycle in phase space is also small. 

Comparing  equation (\ref{eq4}) with the first two equations in (\ref{eq3}), we take as a model for 
the interaction between the Huygens's clocks the system of equations, 
\begin{equation}
\begin{array}{lcl}
 m_1 \ell_1 \ddot \theta_1  + f (\theta_1;\lambda_1, {\tilde \theta}_1)\dot \theta_1  + m_1 g\theta_1  &=&  - m_1 \ddot x_1  \\ 
 m_2 \ell_2 \ddot \theta_2  + f (\theta_2;\lambda_2, {\tilde \theta}_2)\dot \theta_2 + m_2 g\theta_2  &=&  - m_2 \ddot x_2  \\ 
 M\ddot x_1  - f (\theta_1;\lambda_1, {\tilde \theta}_1 ) \dot \theta_1 + 2\rho \dot x_1  - m_1 g\theta_1  &=& k(x_2  - x_1 ) \\ 
 M\ddot x_2  - f (\theta_2;\lambda_2, {\tilde \theta}_2 )\dot \theta_2  + 2\rho \dot x_2  - m_2 g\theta_2  &=&  - k(x_2  - x_1 ) \\ 
 \end{array}
\label{eq6}
\end{equation}
which, by (\ref{eq5}), is a piecewise linear system of equations in an eight-dimensional phase space.

\begin{figure}
 \includegraphics[width= \hsize]{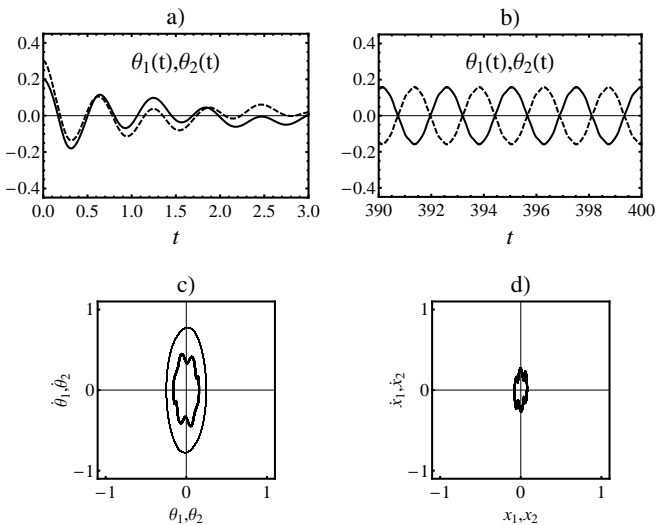}
 \caption{Numerical solutions of the system of equations (\ref{eq6}), with damping function (\ref{eq5}), describing the coupling of two identical pendulum clocks with masses $m=1$ and lengths $\ell=1$. The parameters of the damping function are $\lambda=0.1$ and $\tilde \theta= 0.1$. The  parameter associated with the coupling mechanism of the two-pendulum clocks are: $M=0.1$, $k=10$ and $\rho=0.2$. The initial conditions  are: $\theta_1(0)=0.2$, $\theta_2(0)=0.3$,
$x_1(0)=0$, $x_2(0)=0$, $\dot \theta_1(0)=0$, $\dot \theta_2(0)=0$, $\dot x_1(0)=0$ and $\dot x_2(0)=0$. 
In a) and b), we show the time evolution of the angular coordinates of the two pendulum clocks, before and after anti-phase synchronization, respectively.
In c) and d), we show the asymptotic solutions in a reduced phase space  of the two pendulum clocks  (c), and of the two attachment points  (d). The asymptotic solutions
of the pairs of coordinates $(\theta_1 (t),\dot \theta_1 (t))$ and $(\theta_2 (t),\dot \theta_2 (t))$ coincide. 
In this simulation,  the period of oscillations is $T=2.461$.
For comparison, in c), we show the limit cycle solution (thin line curve) of the reference equation (\ref{eq4}), with damping function (\ref{eq5}). 
}
\label{fig2} 
\end{figure}

To simplify further, we analyze the solutions of  equations (\ref{eq6}), for the particular parameter values, $m_1=m_2=m$,  $\ell_1=\ell_2=\ell$,
$\lambda_1=\lambda_2=\lambda$ and ${\tilde \theta}_1={\tilde \theta}_2={\tilde \theta}$, with  $M>0$. 

In a small neighborhood of the origin in phase space, we can add the first two equations in (\ref{eq6}), and also the third and forth
equations in (\ref{eq6}), and we obtain,
\begin{equation}
\begin{array}{lcl}
 m \ell \ddot \theta -2\lambda \dot \theta  + m g\theta  &=&  - m\ddot x  \\ 
 M\ddot x  + 2\lambda \dot \theta + 2\rho \dot x  - m g\theta  &=& 0 
 \end{array}
\label{eq7}
\end{equation}
where, $\theta=\theta_1+\theta_2$, and $x=x_1+x_2$. Clearly, the solutions of the system of equations (\ref{eq6}) are related with the solutions
of (\ref{eq7}), provided,
$|\theta_1(t)|<{\tilde \theta}$  and $|\theta_2(t)|<{\tilde \theta}$.

The  two nonlinear oscillators described by (\ref{eq6}) exactly synchronize in anti-phase if,
asymptotically in time, for every $h>0$, $\lim_{t_n\to \infty}\theta_1(t_n)=-\lim_{t_n\to \infty}\theta_2(t_n)$, where $t_n=t_0+nh$, $n=0,1,\cdots$, and $t_0$ is the initial time. This condition implies that,
\begin{equation}
\lim_{t_n\to \infty}(\theta_1(t_n)+\theta_2(t_n))=
\lim_{t_n\to \infty} \theta(t_n) =0
\label{eq8}
\end{equation}
So, by (\ref{eq8}), if the zero solution of the system of equations (\ref{eq7}) is asymptotically stable,
and the fixed points of the system of equations (\ref{eq6}) is Lyapunov unstable, then the two pendulum clocks synchronize in anti-phase, \cite{9}. These two stability conditions for the zero fixed points of equations (\ref{eq6}) and (\ref{eq7}) are sufficient to ensure exact anti-phase synchronization.

Writing the system of linear equations (\ref{eq7}) as a first order system of differential equations, we obtain, 
\begin{equation}
\left(\begin{array}{c}
\dot \theta \\ \dot \xi \\ \dot x \\ \dot v \end{array} \right)=
\left(\begin{array}{cccc}
0 & 1 & 0 & 0  \\
a& b& 0 & c \\
0 & 0 & 0 & 1  \\
d & e & 0 & -\ell c 
\end{array} \right)
\left(\begin{array}{c}
 \theta \\ \xi \\ x \\  v\end{array} \right)
\label{eq9}
\end{equation}
where, $\dot \theta=\xi$, $\dot x=v$,  $a=-g(1+{m/ M})/\ell$, $b=2\lambda (1/ (\ell m) +1/(\ell M))$,  $c={2\rho/ (\ell M)}$, $d=gm/M$ and $e=-2\lambda/M$. 
If the eigenvalues of the characteristic polynomial of the matrix in (\ref{eq9})  are non positive, then the system of two interacting  pendulums clocks have  
anti-phase synchronous  solutions, provided the fixed points of system (\ref{eq6}) are Lyapunov unstable, \cite{9}.

\begin{figure}
 \includegraphics[width= \hsize]{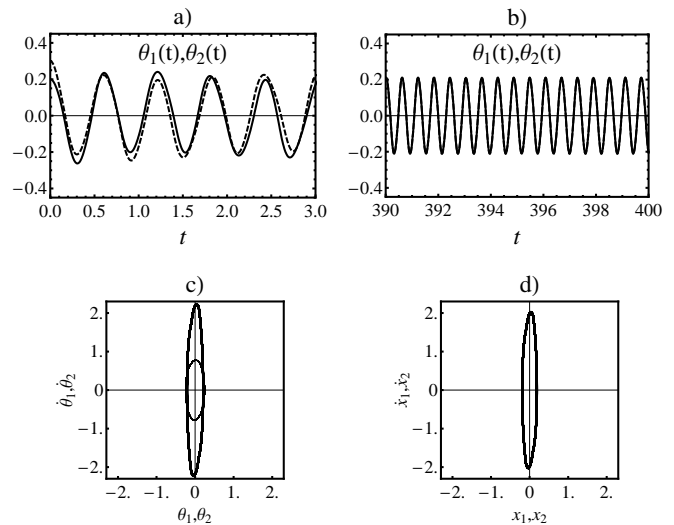}
 \caption{In-phase synchronization of the two pendulum clocks, obtained numerically from the system of equations (\ref{eq6}). The parameters are the same as in Fig.~\ref{fig2}, except the damping parameter $\rho$ that, in this case, has the value $\rho=0.02$.  The initial conditions are the same as in Fig.~\ref{fig2}.  In this simulation, the period of oscillations is $T=0.61$.  In c), the thin line is  the limit cycle solution  of the reference equation (\ref{eq4}), with damping function (\ref{eq5}). }
\label{fig3} 
\end{figure}

The condition of non positivity of the eigenvalues of
the matrix in (\ref{eq9}) can be derived from  the 
the Routh-Hurwitz criterion. It can be shown  (\cite{9}) that the eigenvalues of the characteristic polynomial of the matrix in (\ref{eq9})  are non positive, provided $\rho>\rho_0=(\lambda/\ell) (1 + M/m)$, 
and  $\rho_1 <\rho<\rho_2$,  where  $\rho_1$ and $\rho_2$ are the roots of the polynomial,
\begin{equation}
\begin{array}{rl}
p(\rho)=&4 m \ell \lambda \rho^2  - (4m\lambda^2  + 4M\lambda^2  + 
g m^3\ell )\rho  \\&+ g m^3\lambda   
+  2 g m^2 M\lambda +g m M^2\lambda 
\end{array} 
\label{eq10}
\end{equation}

To analyze numerically the solutions of the system of equations (\ref{eq6}), for the parameters of the  oscillator (\ref{eq4}), we have chosen the parameter values $g=9.8$, $m=1$, $\ell=1$, $\lambda =0.1$, and $\tilde \theta=0.1$. In this case, the uncoupled nonlinear oscillators have the eigen-period 
$T=2.008$.

In Fig.~\ref{fig2}, we show the time evolution of the two pendulum clocks starting  from two different initial amplitudes with zero velocity. The coupling parameters are  $k=10$, $\rho=0.2$ and $M=0.1$. As
$\rho>\rho_0=0.11$, and the roots of the polynomial (\ref{eq10}) are $\rho_1=0.121$ and $\rho_2=24.489$.
The condition of instability of the fixed points of the system (\ref{eq6}) has been calculated numerically and is,
$\rho<0.393$. In the numerical simulations of Fig.~\ref{fig2},
 after a transient time, the exact anti-phase synchronization state is reached. The period of the two pendulum clocks is $T=2.461$,  contrasting with the eigen-period $T=2.008$  of the uncoupled pendulum clocks.
The two pendulum clocks synchronize in anti-phase, in a phase space orbit different from the one obtained if they were uncoupled.  Numerically, the anti-phase synchronized state is an isolated closed  orbit (limit cycle) in the eight-dimensional phase space, and the periods of the angular coordinates $\theta_i$ and of the attachment points $x_i$ are the same. 

\begin{figure}
 \includegraphics[width=0.4 \hsize]{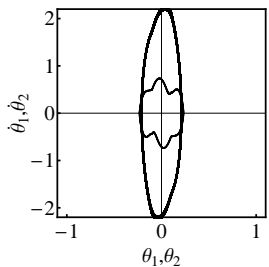}  
 \caption{Stable limit cycles in a reduced phase space of the asymptotic anti-phase and in-phase synchronized states
of the two pendulum clocks, for $\rho=0.02$. The small limit cycle corresponds to  the anti-phase synchronized state, and the larger one to the in-phase synchronized state. The periods of oscillations are $T_{anti}=2.463$, and 
$T_{in}=0.61$. The two stable states are reached changing the initial conditions of the pendulum clocks.}
\label{fig4} 
\end{figure}

Decreasing the  damping parameter $\rho$, the exact anti-phase synchronization regime still persists below the  values,  
$\rho_1=0.121$ and $\rho_0=0.11$. 

Decreasing furthermore the damping parameter $\rho$, and
with the same initial conditions of Fig.~\ref{fig2},
for  $\rho< 0.06$, the two pendulum clocks synchronize with the same phase (in-phase), Fig.~\ref{fig3}.

Changing the initial conditions in the simulations of 
Fig.~\ref{fig3}, from $\theta_1(0)=0.2$ and $\theta_2(0)=0.3$,
to $\theta_1(0)=0.2$ and $\theta_2(0)=-0.3$, we obtain a new asymptotic solution, and the two oscillators still
synchronize in anti-phase. 
For $\rho=0.02$, the system of equations (\ref{eq6}) has two stable limit cycles in the eight-dimensional  phase space, \cite{9}. In Fig.~\ref{fig4}, we show the stable limit cycles associated with the asymptotic anti-phase and the asymptotic in-phase synchronized states
of the two pendulum clocks.

Further numerical analysis shows that there exists a transition region with $\rho$ in the interval $ [0.06,0.07]$, such that,
the anti-phase asymptotic regime corresponds to a limit cycle in phase space, and the  in-phase regime is associated with  a quasi-periodic orbit, \cite{9}. This suggests that the transition between the anti-phase and the in-phase synchronized states is due to a non-local bifurcation, or a sequence of bifurcations.

In conclusion, we have proposed a model describing qualitatively the anti-phase synchronization of clocks as observed by Huygens. The model is consistent 
with the physical mechanism associated with the interactions
between the nonlinear oscillators. The important new issue introduced in the model is the possibility of existence of small movements of the attachment points of the pendulum clocks, a situation
clearly avoid in the modern experimental devices.
This explains why modern experiments have not been able to
reproduce the original Huygen's results.
Anti-phase and in-phase synchrony is obtained with periods
different from the eigen-periods of the individual oscillators. This shows that the equality between the eigen-periods of the individual oscillators is not required to obtain the anti-phase synchronization.
Dropping the small amplitude assumption and  changing  the parameters of the individual oscillators, the results presented here are still true, \cite{9}.

\acknowledgements{ 
This work has been partially supported by a Funda\c c\~ao para a Ci\^encia e a Tecnologia (FCT) pluriannual funding grant to the NonLinear Dynamics Group (GDNL).
}


\begin{references}
\bibitem{9} R. Dil\~ao, On the problem of synchronization of identical dynamical systems:
The Huygens's clocks. To be published.
\end{references}
\end{document}